\documentclass[namedreferences]{solarphysics}
\usepackage[hyperref,optionalrh,solaromanenum]{spr-sola-addons} 
\usepackage{graphicx}                    
\usepackage{color}                       
\usepackage{breakurl}                         

\renewcommand{\vec}[1]{{\mathbfit #1}}

\chardef\us=`\_
\begin{document}

\begin{article}

\begin{opening}
\title{The Formation of a Small-scale Filament after Flux Emergence on the Quiet Sun}
%
\author[addressref={1,2,3},corref,email={chc@ynao.ac.cn}]{\inits{Hechao}\fnm{Hechao}~\lnm{Chen},\orcid{https://orcid.org/0000-0001-7866-4358}}
\author[addressref={1,2}]{\inits{Jiayan}\fnm{Jiayan}~\lnm{Yang},\orcid{https://orcid.org/0000-0003-3462-4340}}
\author[addressref={1,2}]{\inits{Bo}\fnm{Bo}~\lnm{Yang}}
\author[addressref={1,2}]{\inits{Kaifan}\fnm{Kaifan}~\lnm{Ji},\orcid{https://orcid.org/0000-0001-8950-3875}}
\author[addressref={1,2,3}]{\inits{Yi}\fnm{Yi}~\lnm{Bi},\orcid{https://orcid.org/0000-0002-5302-3404}}
%
\runningauthor{H. C. Chen $et$ $al.$}
\runningtitle{Filament Formation}
\address[id={1}]{Yunnan Observatories, Chinese Academy of Sciences,
396 Yangfangwang, Guandu District,Kunming, 650216, China}
\address[id={2}]{Center for Astronomical Mega-Science, Chinese Academy of Sciences, 20A Datun Road, Chaoyang District,Beijing, 100012, China}
\address[id={3}]{University of Chinese Academy of Sciences,19A Yuquan Road, Shijingshan District,Beijing 100049, China}
\begin{abstract}
We present observations of the formation process of a small-scale filament on the quiet Sun during 5-6 February 2016 and investigate its formation cause. Initially, a small dipole emerged and its associated arch filament system was found to reconnect with overlying coronal fields accompanied by numerous EUV bright points. When bright points faded out, many elongated dark threads formed bridging the positive magnetic element of dipole and external negative network fields. Interestingly, an anti-clockwise photospheric rotational motion (PRM) set in within the positive endpoint region of newborn dark threads following the flux emergence and lasted for more than 10 hours. Under the drive of the PRM, these dispersive dark threads gradually aligned along the north-south direction and finally coalesced into an inverse S-shaped filament. Consistent with the dextral chirality of the filament, magnetic helicity calculations show that an amount of negative helicity was persistently injected from the rotational positive magnetic element and accumulated during the formation of the filament. These observations suggest that twisted emerging fields may lead to the formation of the filament $via$ reconnection with pre-existing fields and release of its inner magnetic twist. The persistent PRM might trace a covert twist relaxation from below photosphere to the low corona.
\end{abstract}
%
\keywords{Prominences, Formation and Evolution; Magnetic fields, Photosphere; Velocity Fields, Photosphere; Helicity, Magnetic}
\end{opening}

\section{Introduction}
\label{S-Introduction}
Solar filaments, also known as prominences, are common but complicated structures in the solar atmosphere. They are good indicators of magnetic complexity on the solar surface, and their disruptions are often associated with solar flares and coronal mass ejections (CMEs) \citep{2003ApJ...597L.161J,2012ApJ...749...12Y,2011RAA....11..594S}. Filaments are typically observed to form in filament channels along polarity inversion lines (PILs) and follow a magnetic pattern of chirality \citep{1998SoPh..182..107M}. Through decades of observations, it is recognized that when plenty of cool dense plasma stably accumulated inside correct magnetic structures, filaments will naturally form \citep{2010SSRv..151..333M}. But, the origins of filament magnetic structures and their plasma remain debatable until now.
\par
In general, the magnetic structures of filaments have been explained as a flux-rope configuration or a sheared-arcade configuration, in which magnetic dips were believed to provide upward support against gravity \citep{2010ApJ...714..343G,2014ApJ...784...50C,2016ApJ...830...16Y,2017ApJ...835...94O}. Although no solid observational evidences showed that one of them outdoes the other, the twisted flux rope had been widely adopted in numerous formation models of solar filament/prominence.
Theoretically, flux ropes can be formed $via$ two ways: successive reconnection of sheared arcades in the low corona \citep{1989ApJ...343..971V,2001ApJ...558..872M}, or bodily emergence of a coherent twisted flux rope from the convection zone \citep{1994SoPh..155...69R,1995ApJ...443..818L}.
In the reconnection model, successive reconnection and photospheric flows along the PIL were thought to be two key elements for the creation of a twisted flux rope.
In the emergence model, a twisted flux rope is assumed to be generated in the convection zone (hereafter the primordial flux rope) and then directly emerged into the low corona by buoyancy. Accordingly, cool dense plasma may be lifted up with rising twisted field lines leading to the appearance of a filament \citep{1994SoPh..155...69R,1997SoPh..174...91L}.
\par
To date, many observational and simulation studies have been presented to support the reconnection model \citep{1997ApJ...479..448G,2007ApJ...666.1284W,2016ApJ...816...41Y,2017ApJ...840L..23X,2017ApJ...839..128W}. In particular, \citet{2016ApJ...816...41Y} recently found that a filament even can be rapidly formed $via$ reconnection within around 20 minutes. Referring to such rapid formation of filaments, \citet{2017ApJ...845...12K} conducted 3D MHD simulations under thermal conductions and radiative cooling condition, and they found reconnection can lead not only to flux rope formation but also to filament plasma formation $via$ radiative condensation.
On the other hand, clear observational evidence in favor of the emergence model was quite sparse \citep{1997SoPh..174...91L,2017ApJ...845...18Y}. \citet{2009ApJ...697..913O} investigated the magnetic field evolution below a filament in NOAA active region 10953, and they found the abutting regions along the PIL, which mainly contained horizontal magnetic fields, first grew in size and then became narrowed with time; meanwhile, the orientations of horizontal fields gradually reversed following obvious blueshift. Thus authors stated that an emerging flux rope was observed in their work, and suggested that the emergence of the flux rope contribute to the formation and maintenance of the filament. \citet{2012SoPh..278...33V} revisited the same event but argued that the same observations can also be well explained by magnetic flux cancellation.
\par
Besides, most numerical simulations of emerging primordial flux ropes fail to bodily lift its axis and cool plasma into the corona only using buoyancy and magnetic buoyancy instabilities \citep{2001ApJ...554L.111F,2004ApJ...610..588M,2006ApJ...653.1499M}. To resolve this problem, several possible sub-processes were involved in subsequent simulations, such as magnetic reconnection and helicity transport. \citet{2008A&A...492L..35A} have shown that after the top section of the primordial flux rope has emerged, magnetic reconnection can reconfigure the emerged coronal arcades to produce a secondary flux rope and lift cool plasma to coronal heights. Instead, \citet{2009ApJ...697.1529F} proposed that after the top portion of primordial flux rope emerges from the photosphere, prominent rotational motion may set in within each polarity and further twists up its expanded corona portion to form a newborn flux rope in the corona. They pointed out that photospheric rotational motions (PRMs) of two polarity flux are a manifestation of nonlinear torsional Alfv\'{e}n waves propagating along the field lines. Although these simulations provide possible predictions for the formation of emergence-related filaments, there are few observations to examine their interpretation. Therefore, more new observational work on the formation of filaments over a wide range of latitudes is required.
\par
In the quiet Sun, numerous small-scale filaments unceasingly form and erupt in the polarity-mixed region. Because evidence strongly illustrates that they are miniature counterparts of large-scale filaments \citep{2011ApJ...738L..20H,2017ApJ...835...35H,2012ApJ...745..164S,2017ApJ...851...67S,2017ApJ...844..131P}, we may get a different insight into the formation mechanism of large-scale filaments $via$ investigating the formation processes of these small-scale filaments. In this paper, we report the formation of a small-scale filament in the quiet Sun on 5-6 February 2016. In particular, the filament formation is found to closely relate to a rotational magnetic dipole and its emerging arch filament system (AFS). Based on imaging observations and magnetic helicity calculations, we carefully study the formation mechanism of the filament. Our data sets and methods are presented in the next section. Detail investigation of the filament formation and associated magnetic evolution are shown in Section 3. In Section 4, we summarize and discuss the results.
\section{Observations and Methods}
\label{S-Observations}
\subsection{Observational Data}
Our primary data were obtained from the $Solar$ $Dynamics$ $Observatory$ (SDO), on which the $Atmospheric$ $Imaging$ $Assembly$ \citep{2012SoPh..275...17L} uninterruptedly observes the full-disk of the Sun with a spatial resolution and cadence of 1.$^{\prime \prime}$5 and 12s, respectively, and the $Helioseismic$ $and$ $Magnetic$ $Imager$ \citep{2012SoPh..275..229S} also measures the photospheric magnetic fields at 6173 \AA \ with a spatial sampling of 0$^{\prime \prime}$.5 pixel$^{-1}$. In this work, we focus on EUV 304 \AA \ , 193 \AA \ and 171 \AA \ images to study the formation process of the filament, and utilize the light-of-sight (LOS) magnetogram with a cadence of 720 seconds from the HMI to study its underlying photospheric magnetic field evolution. Meanwhile, the H$\alpha$ line center images from the $Global$ $Oscillation$ $Network$ $Group$ (GONG) instruments \citep{2011SPD....42.1745H}, which have a pixel size of 1.$^{\prime \prime}$05 and cadence of 1 minute, are also complemented.
In addition, the full-disk Disambiguated vector magnetograms with a cadence of 720 seconds from the HMI were utilized to derive the flux transport velocity $via$ the Differential Affine Velocity Estimator for Vector Magnetograms (DAVE4VM) \citep{2008ApJ...683.1134S} method. The full-disk Disambiguated vector magnetograms are inverted $via$ Very Fast Inversion of the Stokes Vector (VFISV) method \citep{2011SoPh..273..267B}, and the 180$^{\circ}$ azimuth ambiguity was resolved by the Minimum Energy method \citep{2006SoPh..237..267M}. In this work, vector magnetic field data were preprocessed by SSWIDL modules of hmi$_{-}$disambig.pro and hmi$_{-}$b2ptr.pro. More detailed information on vector magnetic field data processing can be found in \citet{2014SoPh..289.3483H} and \citet{2012ApJ...748...77S}. Note that this event occurred near the solar disk-center (around N15$^{\circ}$W15$^{\circ}$), and projection effects thus can be ignored. To compensate for solar rotation, all the multi-wavelength data taken at different times are aligned to an appropriate reference time (16:00 UT on 2016 February 6).
\subsection{Magnetic Helicity Injection and Flux Density Distribution}
Magnetic helicity is a helpful metric to describe the magnetic complexity within a finite volume. In the Sun, magnetic helicity either arises from emerging twisted flux from below photosphere (emergence term, ${\rm d}H_{\rm e}/{\rm dt}$), or is generated $via$ tangential photospheric motions on the solar surface (shear term, ${\rm d}H_{\rm s}/{\rm dt}$). Thus, the helicity flux across a planar surface $S$ is defined as \citep{1984JFM...147..133B}
\begin{equation}
\frac{{\rm d}H}{{\rm dt}}\Biggr |_{S}=\underbrace{2\int_{\rm s}(\vec{A}_{{\rm p}}\cdot \vec{B}_{{\rm t}})V_{{\rm n}}{\rm d}S}_{{\rm d}H_{\rm e}/{\rm dt}}-\underbrace{2\int_{s}(\vec{A}_{{\rm p}}\cdot \vec{V}_{{\rm t}})B_{{\rm n}}{\rm d}S}_{{\rm d}H_{\rm s}/{\rm dt}},
\end{equation}
where $\vec{A}_{{\rm p}}$ is the vector potential of the potential field $\vec{B}_{{\rm p}}$, $\vec{B}_{{\rm t}}$ and $B_{\rm n}$ are the normal and tangential magnetic fields, and $\vec{V}_{\rm n}$ and $\vec{V}_{\rm t}$ are the tangential and normal components of $\vec{V}$, the plasma velocity. \citet{2003SoPh..215..203D} showed that the plasma velocity and magnetic fields at the photosphere can be combined to derive the flux transport velocity, $\vec{u}$, i.e. the apparent horizontal footpoint velocity of field lines at the photosphere:
\begin{equation}
\vec{u}=\vec{V}_{\rm t}-\frac{V_{\rm n}}{B_{\rm n}}\vec{B}_{\rm t}.
\end{equation}
Based on Equation 1 and 2, \citet{2005A&A...439.1191P} suggested helicity flux can also be written as
\begin{equation}
\frac{{\rm d}H}{\rm dt}\Biggr |_{S}=\frac{-1}{2 \pi}\int_{S}\int_{S^{\prime}} {\frac{[(\vec{x}-\vec{x^{\prime})}\times(\vec{u}-\vec{u^{\prime}})]|_{\rm n}}{|\vec{x}-\vec{x{^{\prime}}}|^2}}B_{{\rm n}}(\vec{x}) B_{{\rm n}}(\vec{x^{\prime}}){\rm d}S{\rm d}S^{\prime},
\end{equation}
where $\vec{x}$ and $\vec{x^\prime}$ indicate the photospheric footpoints of magnetic flux tubes.
\par
It is worth noting that the field-aligned component of plasma velocity $\vec{V}_{\|}$ corresponds to the irrelevant field-aligned plasma flow velocity, which needs to be removed in the computation of helicity flux \citep{2012ApJ...761..105L}. In other words, Equation 2 need be corrected as
\begin{equation}
\vec{u}=\vec{V}_{\bot {\rm t}}-\frac{{V_{\bot {\rm n}}}}{{B_{\rm n}}}\vec{B_{\rm t}},
\end{equation}
where $\vec{V}_{\bot}=\vec{V}-(\vec{V}\cdot \vec{B}/B^{2})\vec{B}$, and $\vec{V}_{\bot {\rm t}}$ and $V_{\bot {\rm n}}$ are the tangential and normal components of the velocity $\vec{V}_{\bot}$. Meanwhile, the shear and emergence terms of Equation 1 can be rewritten as
\begin{equation}
\frac{{\rm d}H_{\rm e}}{\rm dt}=\frac{1}{2 \pi}\int_{S}\int_{S^{\prime}} {\frac{\vec{x}-\vec{x^{\prime}}}{|\vec{x}-\vec{x{^{\prime}}}|^2}}{\rm d}S{\rm d}S^{\prime}
\times[\vec{B}_{\rm t}(\vec{x})V_{\bot {\rm n}}(\vec{x})B_{\rm n}(\vec{x^{\prime}})-\vec{B}_{\rm t}(\vec{x^{\prime}})V_{\bot {\rm n}}(\vec{x^{\prime}})B_{\rm n}(\vec{x})],
\end{equation}
\begin{equation}
\frac{{\rm d}H_{\rm s}}{\rm dt}=\frac{-1}{2 \pi}\int_{S}\int_{S^{\prime}} {\frac{\vec{x}-\vec{x^{\prime}}}{|\vec{x}-\vec{x{^{\prime}}}|^2}}{\rm d}S{\rm d}S^{\prime}\times[(\vec{V}_{\bot {\rm t}}(\vec{x})-\vec{V}_{\bot {\rm t}}(\vec{x^{\prime}}))B_{\rm n}(\vec{x})B_{\rm n}({\vec{x^{\prime}})}],
\end{equation}
(see also Liu $et$ $al.$, 2014; Bi, $et$ $al.$, 2016; Vemareddy and D$\acute{e}$moulin, 2017).
Accordingly, the helicity flux density distribution $G_\theta$ can be derived by
\begin{equation}
\int_{S} G_{\theta} {\rm d}S=\frac{{\rm d}H}{\rm dt}\Biggr |_{S}=\frac{{\rm d}H_{ \rm e}}{\rm dt}+\frac{{\rm d}H_{\rm s}}{\rm dt}.
\end{equation}
Since magnetic helicity is not a local quantity, the distribution of helicity flux is only meaningful when one considers a whole elementary flux tube rather than its footpoints individually \citep{2005A&A...439.1191P,2014SoPh..289..107D}. By taking the magnetic connectivity into account, \citet{2005A&A...439.1191P} introduced an improved definition of a surface-density of helicity flux, $G_{\phi}$,
\begin{equation}
G_\phi (\vec{x}_{\rm a\pm})=\frac{1}{2} \Biggr\{ G_\theta(\vec{x}_{\rm a\pm})+G_\theta(\vec{x}_{\rm a\mp}) \biggr |{\frac{B_{\rm n}(\vec{x}_{\rm a\pm})}{B_{\rm n}(\vec{x}_{\rm a\mp})} \biggr |}\Biggr\},
\end{equation}
which is a redistribution of $G_{\theta}$ at both footpoints of an elementary flux tube.
Here, $a$ indicates a closed elementary flux tube which is anchored at $\vec{x}_{\rm a \pm}$ in the photosphere.
\par
In this study, the plasma velocity $\vec{V}$ derived from the DAVE4VM method \citep{2008ApJ...683.1134S} is used to compute the flux transport velocity, $\vec{u}_{\rm dave4vm}$. Based on $\vec{u}_{\rm dave4vm}$, we compute the helicity flux density and the total helicity flux, and further compare with the results obtained when the flux transport velocity, i.e. $\vec{u}_{\rm dave}$, is computed from Differential Affine Velocity Estimator (DAVE) method \citep{2006ApJ...646.1358S}. The computed window size is 19$\times$19 pixels in each case. Then we computed helicity flux and accumulation with Equation 3 and Equation 5-6, respectively. Moreover, to check the spatial distribution of helicity injection rate in the region of filament activity, we also calculated the connectivity-based helicity flux density distribution $G_{\phi}$ with Equation 7 and 8 \citep{2013A&A...555L...6D,2015ApJ...805...48B}. The connectivity information over the computed region is obtained from a linear force free field (LFFF) extrapolation \citep{1981A&A...100..197A,1989ApJS...69..323G}. The force-free parameter value, $\alpha$, in the LFFF extrapolation is approximatively determined from the comparison between magnetic dips and the actual filament.
\begin{figure}    
   \centerline{\includegraphics[width=1.2\textwidth,clip=]{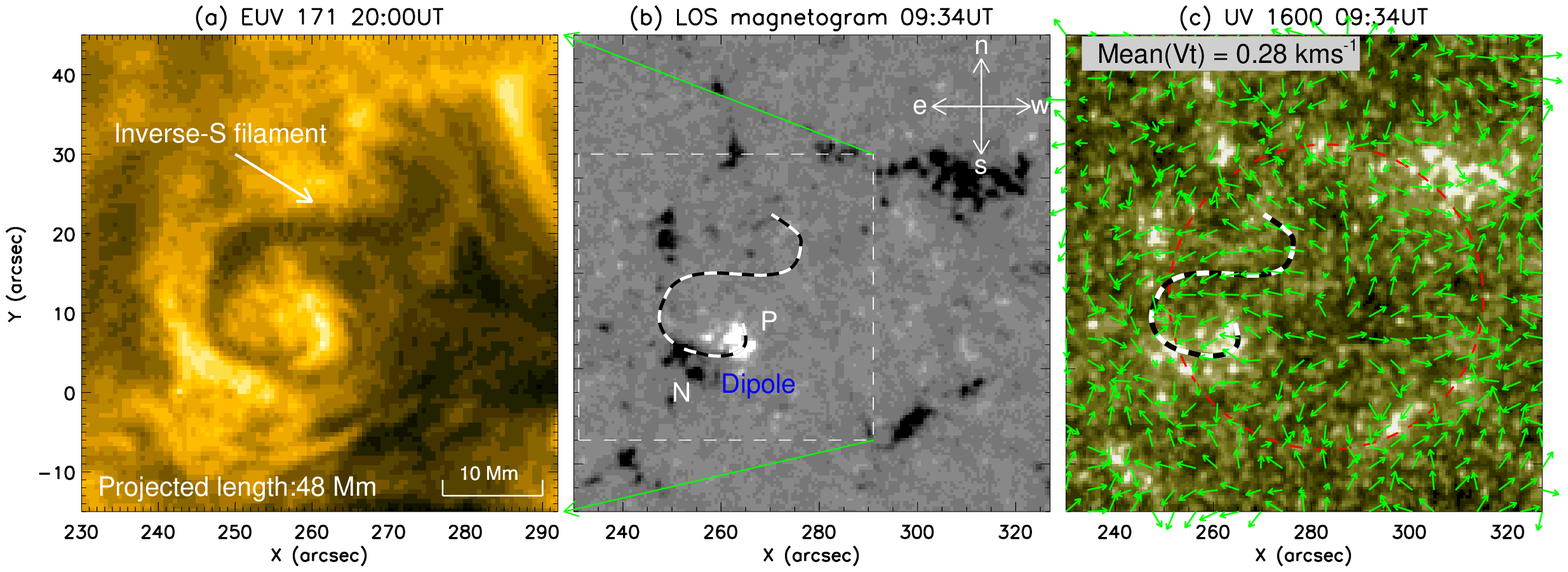}
              }
              \caption{AIA 171 \AA \ (a), HMI line-of-sight magnetogram (b) and 1600 \AA \ (c) images showing the formation environment of the filament. The dashed box in panel (b) indicates the field of view (FOV) of panel (a). Magnetic flux P/N denotes the positive/negative magnetic polarity of the dipole, which are saturated at $\pm$ 100G, respectively.
              The black-and-white curve lines outline the axis of the filament. The green arrows in panel (c) donate the tangential plasma velocity field derived from derived from 12 min cadence vector magnetograms $via$ the DAVE4VM method, which was averaged over a one hour time period.} The red circle circumscribes the supergranular cell whose mean velocity was around 0.28 km s$^{-1}$.
   \label{F-1}
   \end{figure}
\section{Results}
\label{S-Results}
\begin{figure}    
   \centerline{\includegraphics[width=1.0\textwidth,clip=]{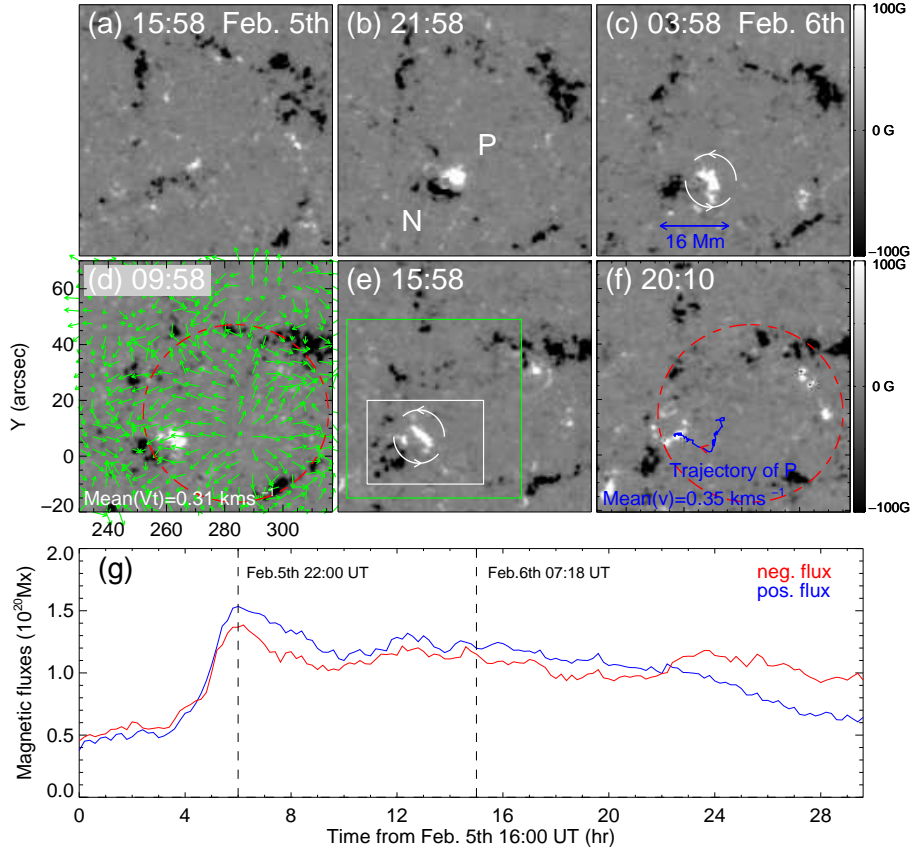}
              }
              \caption{(a-f): The photospheric evolution over the region of filament activity (see animation hmi.mov in the online version). P and N denote the positive and negative polarity of the dipole. White circular arrows represent the anticlockwise rotating motion of P. The green arrows in (d) indicate the tangential plasma
              velocity field derived from 12 min cadence vector magnetic fields $via$ DAVE4VM method, which was averaged over a one hour time period. The red circles in (d) and (f) outline the supergranular cell. In (e), green rectangle denotes FOV of Figure 3,5,6 and 8, while white rectangle denotes FOV of Figure 4. (g): The time change of magnetic flux computed within the white rectangle of (e).}
   \label{F-2}
   \end{figure}
Figure 1 presents the general formation environment of the filament. The left image shows that the formed small filament displayed an inverse-S shape with an apparent length of about 48 Mm in the 171 \AA \ channel. In the middle image, a small emerged dipole, which consisted of flux elements P and N surrounded by opposite-polarity network fields, can be clearly seen from the LOS magnetogram of HMI. When the axis of the filament is superimposed on the magnetogram, it is found that the filament anchored its south endpoints at P, while the root of its north endpoints is in negative-polarity intranetwork fields. Since the magnetic component along the axis of the filament points to the right, the filament should be dextral according to the chirality definition of \citet{1998SoPh..182..107M}. In the 1600 \AA \ image (the right one), one can see that the filament is located near the boundary of an obvious chromospheric network. Based on the tangential plasma velocity fields calculated by DAVE4VM, the chromospheric network was found to co-located with a supergranular cell (indicated by the red circle). In other words, the filament actually formed within a supergranular cell.
\par
\begin{figure}    
   \centerline{\includegraphics[width=1.0\textwidth,clip=]{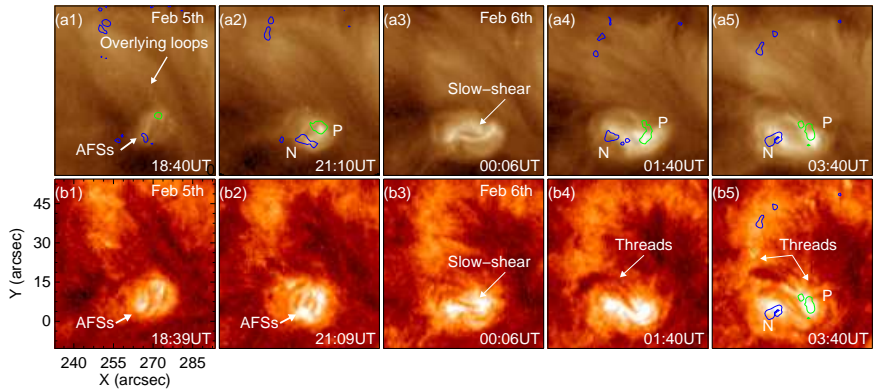}
              }
              \caption{The slow rising of arch filament system (AFS) on 193 \AA \ (the top raw) and 304 \AA \ (the bottom raw) images. Green/blue contours in the top raw denote positive/negative magnetic polarity (P/N) of the dipole, which are saturated at $\pm$ 100 G.}
   \label{F-3}
   \end{figure}
Compared with the calm negative-polarity fields near the north endpoints of the filament, the magnetic fields in its south endpoint region underwent conspicuous changes. A small dipole emerged in the supergranular cell around 16:00 UT on 5 February (Figure 2 a-c).
Soon the opposite-polarity fluxes, P and N, rapidly separated from each other with maximum flux up to 10$^{20}$ Mx (see Figure 2g). At the same time, an AFS arose in AIA 304\AA \ and 171\AA \ channels with cool plasma trapping in their arch-shaped field lines \citep{1967SoPh....2..451B}.
Initially, the plasma-trapped fibrils of AFS crossed P and N nearly perpendicularly (see the bottom raw of Figure 3). Then they gradually became sheared due to a slow shear between P and N, implying the emerging dipole possessed twisted fields. Note that there existed a series of overlying coronal loops above the AFS (see the top row of Figure 3). With the on-going separation of P and N, the AFS continually ascended and further came into contact with overlying loops. Due to the interplay between the AFS and overlying loops, some elongated chromospheric threads were clearly created by 03:40 UT connecting P and peripheral negative-polarity fields.

\begin{figure}    
   \centerline{\includegraphics[width=1.0\textwidth,clip=]{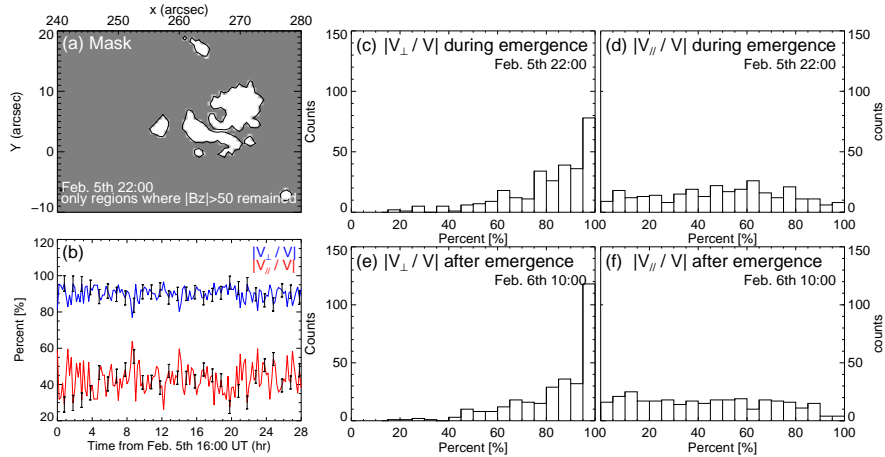}
              }
              \caption{(a): The computed mask defined to eliminate background effects. The computed FOV: $x$=[240,280] arcsec, $y$=[-10,20] arcsec. With this mask, the velocity information within regions satisfying abs$(\vec{B}_{z})>50$ G is extracted for computation of $|V_{\bot}/V|$ and $|V_{\|}/V|$ above the whole dipole. Black contours in panel (a) denote the outline of the dipole, which are saturated at $\pm$ 50 G. (b): The temporal evolution of spatial median of $|V_{\bot}/V|$ (blue) and $|V_{\|}/V|$ (red) for the dipole. The 1$\sigma$ errors are shown by black bars. (c-d) and (e-f): Histograms of $|V_{\bot}/V|$ and $|V_{\|}/V|$ for the dipole obtained during and after its emergence, respectively. The size of each bin in histograms denotes 5 $\%$.}
   \label{F-4}
   \end{figure}
\par
When the separation of P and N ceased, the dipole subsequently drifted towards the east boundary of the supergranular cell (see Figure 2 d-f). The trajectory of P was carefully traced and plotted by the blue curve in Figure 2f. The mean migration velocity of P was calculated as 0.35 km s$^{-1}$, which is roughly consistent with the mean value of the calculated tangential velocity $\vec{V}_{\rm t}$ (around 0.29--0.32 km s$^{-1}$). Moreover, P migrated towards the supergranular boundary along a curved trajectory, indicating co-existence of a radial and a circular component of velocity inside the supergranular flow \citep{1998A&A...335..341Z}. The most interesting thing is that P was found to demonstrate obvious anti-clockwise PRM and lasted for more than 10 hours after its emergence (as marked by circular arrows in Figure 2c and 2e, and also see the online animation, hmi.mov).
To rule out the possibility that the PRM might simply be the apparent phenomenon of plasma draining along the field-lines of the emerging twisted magnetic fields, we inspect the normal and tangential components of plasma velocity, $\vec{V}_{\bot}$ and $\vec{V}_{\|}$, within a zoomed FOV (denoted by the white rectangle in Figure 2e). To eliminate background effects, a mask was defined to extract the velocity information within regions where satisfying abs($\vec{B}_{z}$)$\geq$ 50 gauss (see Figure 4a). With the mask, the unsigned values of $|V_{\bot}/V|$ and $|V_{\|}/V|$ were measured over the dipolar region with time.
Figure 4b illustrates the temporal evolution of the spatial median of $|V_{\bot}/V|$ and $|V_{\|}/V|$ computed above the dipole. The evolutionary characteristics of $|V_{\bot}/V|$ and $|V_{\|}/V|$ are well above the uncertainties that are estimated by the root mean square of 50 Monte Carlo experiments. In each experiment, a Gaussian noise was added to three components of the vector magnetic field. The width of the Gaussian function is 100 G, which is roughly the photon noise level of the vector magnetic field \citep{2014SoPh..289.3483H}. Note that to better show the result curves, original errors were averaged over a one hour time period and only plotted at several representative instants in Figure 4b.
It is clear that the plasma velocity was mainly normal to the vector magnetic fields throughout the 28 hours after the dipole emerged.
Moreover, such situation could also been clearly identified from histograms of $|V_{\bot}/V|$ and $|V_{\|}/V|$ for the dipole obtained during and after its emergence in Figure 4c-f, respectively.
Therefore, the anti-clockwise PRM should corresponded to a real photospheric motion instead of an apparent phenomenon caused by plasma draining during the flux emergence.
\par
The formation of the filament took place during 03:40 to 20:00 UT on 6 February. For ease of description, we divided the formation of the filament into two forming stages.
Interestingly, the anti-clockwise PRM of P appeared to play a role in both stages.
Figure 5 shows the first stage of the filament formation, in which reconfiguration of chromospheric threads was observed in 304 \AA \ images. To better display the magnetic reconfiguration, we outline the axes of some prominent threads in Figure 5a and Figure 5e, respectively. Initially, these chromospheric threads connected P and peripheral negative fields in the northeast-southwest direction at 03:43 UT. Due to the anticlockwise PRM of P, many chromospheric threads gradually changed their original orientations. Meanwhile, episodes of EUV bright points were found near their negative-polarity ends (circled by green lines in Figure 5b and 5c). As a result, chromospheric threads progressively aligned along the north-south direction at 05:38 UT. In Figure 5f, the brightness curve of  AIA 304\AA \ calculated in the black dashed box is plotted comparing with time. Obviously, during the reconfiguration process, some intermittent peaks appeared in the AIA brightness curve during 04:30-05:00 UT, which indicates that magnetic reconnection might have been involved in this process.
   \begin{figure}    
   \centerline{\includegraphics[width=1.0\textwidth,clip=]{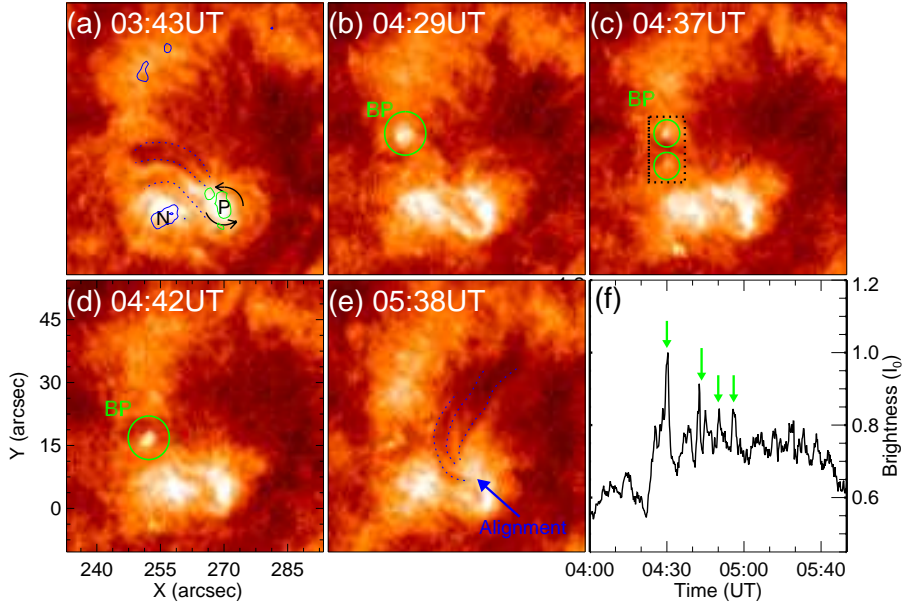}
              }
              \caption{The reconfiguration of chromospheric threads in 304 \AA \ images (see the online animation 171-304.mov). Green/blue contour in panel (a) denotes the positive/negative polarity of the dipole (P/N), which is saturated at $\pm$ 100 G. The black ellipse arrow in panel (a) denotes the anticlockwise rotation of P.
              Green circles denote episodes of EUV bright points (BP). Blue dashed curves in panels (a) and (e) outline the axes of dark threads. Panel (f): normalized brightness of 304 \AA \ images computed in the area denoted by black dashed box in panel (c). Green arrows denote the occurrence of BPs.}
   \label{F-5}
   \end{figure}
\par
The second stage of the filament formation was shown in Figure 6 (also see the online animation, 171-304.mov). Based on GONG H${\alpha}$ observations, we notice that the filament first appeared at 07:18 UT demonstrating a straight shape (Figure 6c2), and then it evolved into an inverse-S shaped filament by 20:00 UT (Figure 6c3). This suggests that more magnetic non-potentiality should be stored during this course. Using EUV 304 \AA \ and 171 \AA \ images, we further inspect the detailed formation process of the inverse-S shaped filament. At 05:40 UT, the aligned chromospheric threads displayed as a compact structure with a south endpoint anchored at P (Figure 6a1 and 6b1). Note that P corresponded to an EUV patch, which also persistently displayed an apparent anticlockwise rotation. Accordingly, the field-connected dark threads were continuously dragged and twisted by the rotating EUV patch. From 05:40 to 10:30 UT, one can see that dark threads gradually got sheared. As the anticlockwise rotation of the EUV patch continued, dark threads progressively coalesced into a distinct filament in both 304 \AA \ and 171 \AA \ images by 15:30 UT (marked by green arrows in Figure 6a4 and 6b4). Several hours later, it is noted that the filament became more distinguishable from its background, and finally developed into a slender inverse-S shape by 20:00 UT. These observations give us a clear clue that the formation of the filament was strongly related to its underlying PRM.
\par
      \begin{figure}    
   \centerline{\includegraphics[width=1.4\textwidth,clip=]{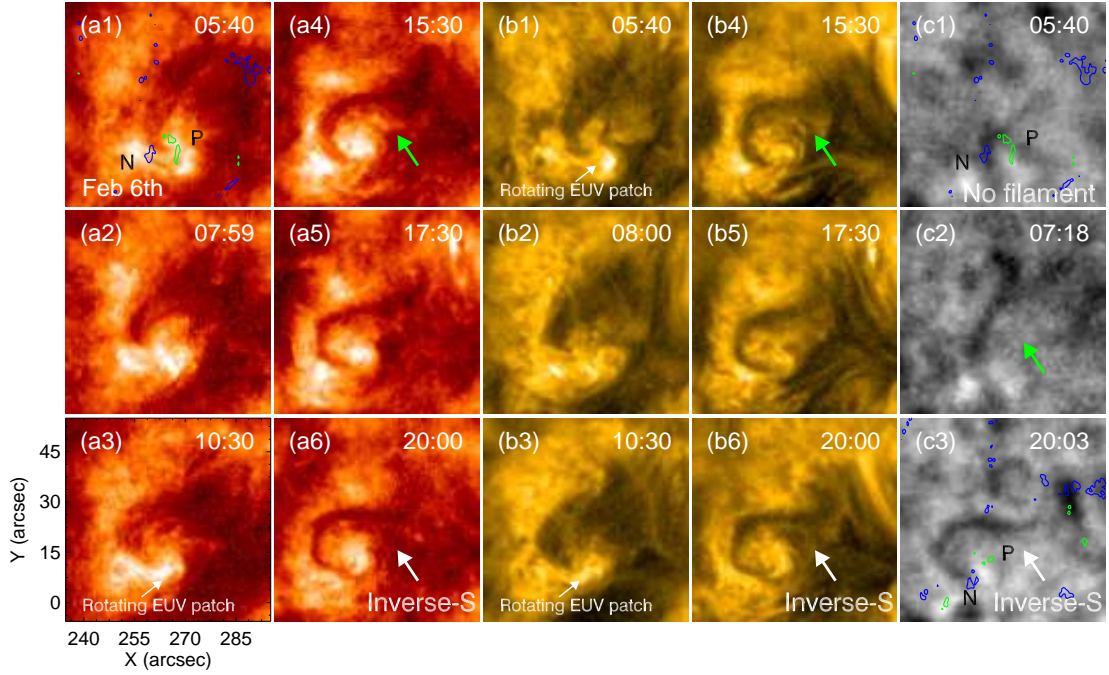}
              }
              \caption{The second forming stage of the filament in AIA 304 \AA \ (a1-a6), 171 \AA \ (b1-b6) and GONG H$_{\alpha}$ (c1-c3) images (see the online animation 171-304.mov). Green/blue contours in panels (a1), (c1) and (c3) denote positive/negative magnetic polarities (P/N) of the dipole, which are saturated in $\pm$ 100 G. Green arrows in panels (a4), (b4) and (c2) denote the forming filament.}
   \label{F-6}
   \end{figure}
      \begin{figure}    
   \centerline{\includegraphics[width=1.4\textwidth,clip=]{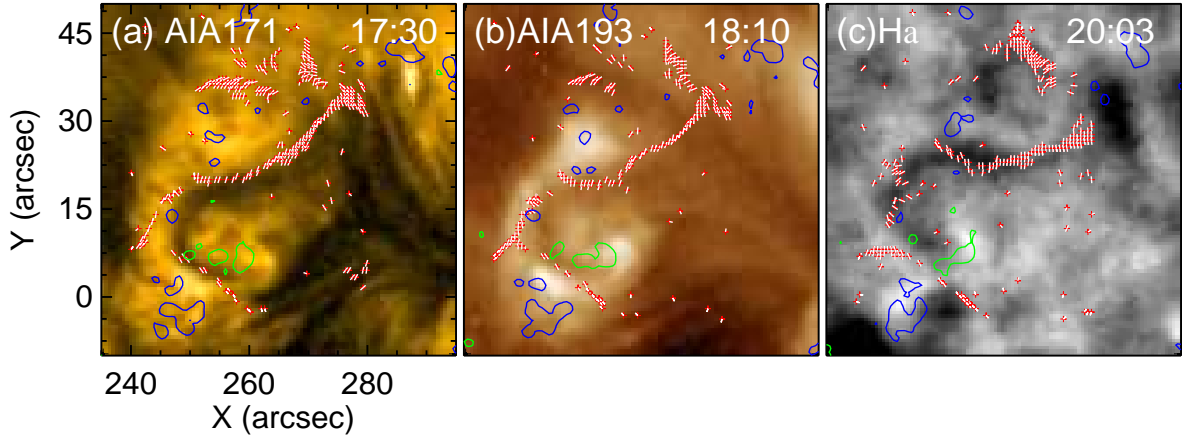}
              }
              \caption{The comparison between the actual filament and magnetic dips obtained from linear force-free extrapolations with $\alpha=-0.015$ Mm$^{-1}$. White field lines are traced from dips (red + signs).
              Green/blue contours denote positive/negative magnetic polarities (P/N) of the dipole, which are saturated in $\pm$ 80 G.}
   \label{F-7}
   \end{figure}
\par
In general, magnetic dips along the magnetic field lines are defined by $B_{z}=0$ and ${\vec{B}\cdot\nabla \vec{B}_{z}> 0}$ \citep{2002ApJ...567L..97A,2010ApJ...714..343G,2015ApJ...805...48B}. Considering filament plasma are thought to be collected in magnetic dips, thus the best-fit value of $\alpha$ for the linear force-free extrapolations can be approximately determined from the comparison between the magnetic dips and the actual filament. As presented in Figure 7, the locations of magnetic dips are well in agreement with the location of the filament at different times for $\alpha = 0.015$ Mm$^{-1}$. Taking this value of $\alpha$, we obtained the connectivity information of field lines performing linear force-free extrapolations, and further derived magnetic helicity flux density distributions in the formation course of the filament. In Figure 8, the helicity flux density distributions with $G_{\theta}$ (left two rows) and $G_{\phi}$ maps (right two row) are presented from 6 February 02:00 UT to 20:00 UT.
It is clear that both the $G_{\theta}$ and $G_{\phi}$ maps display mixed signals, but negative signals mainly appear centering around the positive-polarity magnetic element of the dipole, P, where the filament rooted its positive-polarity ends. Compared with the $G_{\theta}$ maps, some weak opposite signals appear in $G_{\phi}$ maps, as found in previous studies \citep{2013A&A...555L...6D,2015ApJ...805...48B}.
In particular, these negative signal lasted for more than 10 hours, implying that persistent negative helicity was injected around the positive-polarity ends of the forming filament.
\par
Figure 9 shows magnetic helicity fluxes and their time accumulation computed over the FOV of Figure 8.
The evolutionary characteristics of these fluxes are well above their uncertainties. Their uncertainties are also estimated by conducting 50 Monte Carlo experiments, the same as for $|V_{\bot}/V|$ and $|V_{\|}/V|$. Similarly, original errors were averaged over a one hour time period and only plotted at several representative instants. In particular, to obtain the uncertainties of helicity flux derived from $\vec{u}_{\rm dave}$, Gaussian noise was added to the LOS magnetic fields in each experiment and the width of its Gaussian function is 10 G.
In Figure 9a-c, both results from $\vec{u}_{\rm dave}$ and $\vec{u}_{\rm dave4vm}$ show that persistent negative helicity was injected from 5 February 22:00 UT to 6 February 21:46 UT.
As a result, a significant amount of negative helicity (around $-8.3\times 10^{38}$ Mx$^{2}$) accumulated over the formation course of the filament.
During this time period, it is found that helicity flux is generated $via$ both the emergence term (around $-2.5\times 10^{38}$ Mx$^{2}$) and the shear term (around $-5.8\times 10^{38}$ Mx$^{2}$), but the contributions from the shear term are more prominent ($\approx$ 80 $\%$)(see Figure 9d-f). These results clearly suggest that the anticlockwise PRM played an important role in injecting negative helicity and left-handed magnetic twist during the formation course of the filament.
In addition, it is worth noting that the total helicity accumulation computed from $\vec{u}_{\rm dave}$ (around $-6.0\times 10^{38}$ Mx$^{2}$) is only comparable to the shear helicity accumulation derived from $\vec{u}_{\rm dave4vm}$. This is in agreement with the studies of \citet{2008ApJ...683.1134S}, \citet{2012ApJ...761..105L}, and \citet{2017A&A...597A.104V} showing that DAVE is weakly sensitive to the ${V}_{\rm n}$ term.
 \begin{figure}    
   \centerline{\includegraphics[width=1.2\textwidth,clip=]{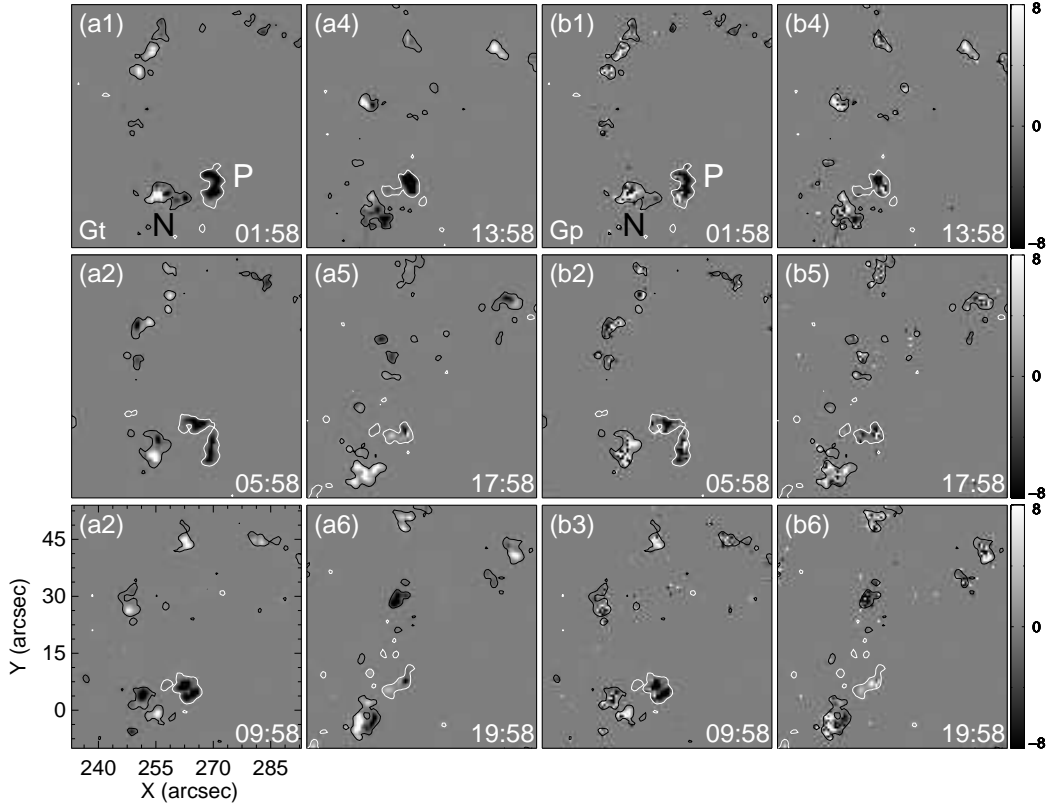}
              }
              \caption{
              Helicity flux distributions with G$_{\theta}$ maps (a1-a6) and G$_{\phi}$ maps (b1-b6) during the formation course of the filament. G$_{\phi}$ is derived from G$_{\theta}$ by taking into account the magnetic connectivity of the field line footpoints.
              White/black contours denote positive/negative magnetic polarities (P/N) of the dipole, which are saturated at $\pm$ 80 G. The computed FOV is $x$=[233,293] arcsec, $y$=[-10,54] arcsec. In the colorbar, helicity flux density distributions are in units of 10$^{21}$ Mx$^2$ m$^{-2}$ s$^{-1}$.
              }
   \label{F-8}
   \end{figure}
    \begin{figure}    
   \centerline{\includegraphics[width=1.2\textwidth,clip=]{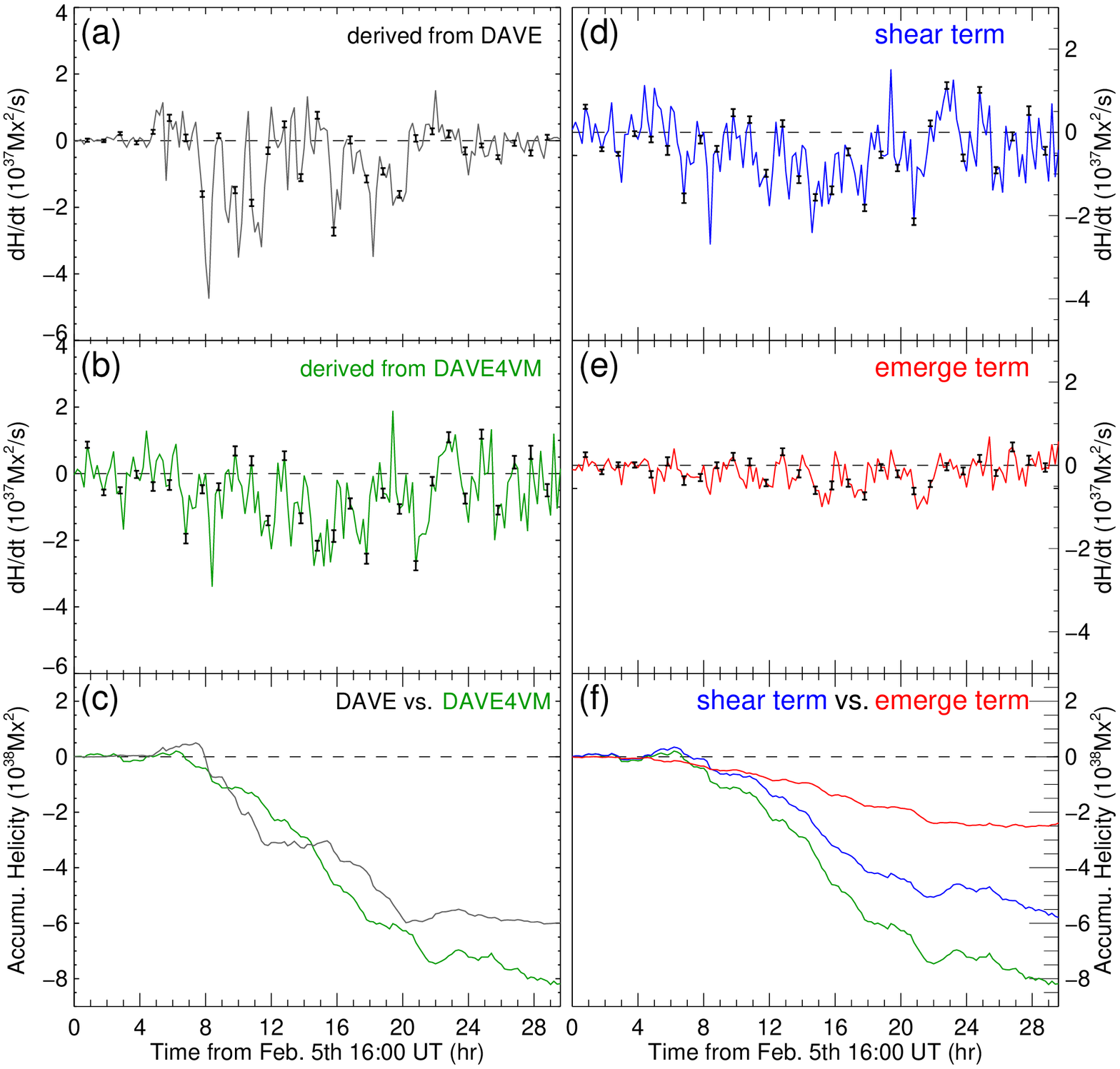}
              }
              \caption{(a): magnetic helicity flux derived from $\vec{u}_{\rm dave}$; (b): magnetic helicity flux derived from $\vec{u}_{\rm dave4vm}$; (c): total magnetic helicity accumulations derived from $\vec{u}_{\rm dave}$ and $\vec{u}_{\rm dave4vm}$; (d): shear helicity flux derived from $\vec{u}_{\rm dave4vm}$; (e): emerge helicity flux derived from $\vec{u}_{\rm dave4vm}$; (f): temporal profiles of magnetic helicity. Blue curve demotes total helicity accumulation, while blue and red curves refer to accumulated helicities from shear and emergence terms. The computed FOV is $x$=[233,293] arcsec, $y$=[-10,54] arcsec. The 1$\sigma$ errors are shown by black bars and only plotted at several representative times.
              }
   \label{F-8}
   \end{figure}
\section{Conclusion and Discussion}
\label{S-Conclusion}
As discussed, formation mechanisms of filament structure remain in dispute. Theoretically, two possible mechanisms (the reconnection model and the emergence model), were proposed to explain the formation of filament magnetic structure. However, compared with the reconnection model, direct supportive observations for the emergence model are still quite rare. In this study, we present a clear formation process of an emergence-related small filament in the quiet Sun with SDO/AIA and HMI observations.
The main analysis results are summarized as follows:
\begin{itemize}
  \item The small-scale filament formed due to a dipolar emergence but only rooted its positive-polarity ends inside the dipole. Moreover, the formed filament (around 48 Mm) was larger than the tiny emerged dipole (around 16 Mm) in spatial scale (see Figure 1a and Figure 2c).
  \item Some plausible evidence for the reconfiguration of magnetic fields, including the creation, alignment and coalescence of chromospheric dark threads, was clearly observed during the formation process of the filament. In particular, numerous EUV bright points intermittently appeared during the alignment and coalescence of dispersive dark threads.
  \item The anticlockwise photospheric rotational motion (PRM) that formed within positive magnetic element of the dipole coincided with the formation of the filament, both spatially and temporally. It is likely that the PRM not only propelled the alignment of dark threads, but also led to the formation of the filament $via$ dragging and twisting the footpoints of dispersive chromospheric threads.
  \item The forming filament gradually changed its apparent shape from straight to a slender inverse-S shape, which might indicated that the magnetic non-potentiality of the filament increased. In accord with the dextral chirality of the filament, our magnetic helicity calculations reveal that an amount of negative helicity was persistently injected from the rotational positive magnetic element of dipole for more than 10 hours.
\end{itemize}
\par
In the emergence model, a primordial flux rope is believed to form below the photosphere. Then due to magnetic buoyancy, the flux rope directly emerges into the low corona, dragging cool plasma with it, lead to the creation of filament within a magnetic dipolar configuration. Such a scenario has been considered in previous MHD numerical simulations, but most of them showed that the emerging primordial flux rope only can partially emerged to the low corona due its trapped cool dense plasma. Later on, some researchers \citep{2004ApJ...610..588M,2008A&A...492L..35A,2009ApJ...697.1529F} suggested that when magnetic reconnection or helicity transport processes are introduced in the expended portion, a secondary flux rope may form above the dipole. In particular, \citet{2009ApJ...697.1529F} proposed that after the primordial flux rope partially emerges into the corona, prominent rotational motion may appear within each polarity due to the buildup of a gradient of the rate of twist along the primordial flux rope. As a result, a secondary flux rope may form in the corona when nonlinear torsional Alfv\'{e}n  waves propagate along field lines, transporting magnetic twist from the inner Sun to the corona.
Here our event not only provides some observational support for these numerical simulations \citep{2004ApJ...610..588M,2008A&A...492L..35A,2009ApJ...697.1529F}, but also reveals several distinct observational features. First, the filament formed due to a dipolar emergence but only rooted its positive-polarity ends inside the dipole. Interestingly, the formed filament had a larger spatial scale than the tiny emerged dipole. Moreover, some plausible evidence for reconnection, such as the occurrence of EUV bright points, and the creation and alignment of chromospheric threads, also was found. These three features strongly imply that new magnetic linkages between emerged fields and pre-existing background fields were established $via$ magnetic reconnection during the formation course of filament. In addition, similar to the simulation results of \citet{2009ApJ...697.1529F}, in our event, persistent anti-clockwise PRM set in within the positive magnetic element of the dipole after flux emergence started and appeared to propel the formation of the filament. Considering the linkages between twisted emerging fields and background potential fields $via$ reconnection, a gradient of the rate of twist along field lines may also be built up. As a result, a similar twist transport process from the inner Sun to the corona may also be initiated during flux emergence. Thus, we suggest that the emerging twisted fields may lead to the formation of the filament $via$ reconnection with pre-existing coronal fields and by release of its inner magnetic twist.
\par
PRMs had been supposed as an attractive magnetic energy and helicity injection source in the solar atmosphere for a long time \citep{1988Natur.335..238B,2011ApJ...741L...7Z,2012ApJ...744...50J,2012ApJ...756L..41S,2012Natur.486..505W}. In active regions, PRMs usually co-locate with rotating sunspots, and many observational works have highlighted their roles in triggering solar eruptions \citep{2014ApJ...784..165R,2016ApJ...829...24V,2017ApJ...836..235L}. In the quiet Sun, PRMs commonly co-locate with ubiquitous short-lived emerged magnetic features \citep{2014ApJ...781....7Y}. Recently, \citet{2015ApJS..219...17Y,2016ApJ...832...23Y} found that rotating sunspots may play an important role in the formation of adjacent active-region filaments. \citet{2015ApJ...803...86Y} also noticed that small-scale rotating network fields can prompt a J-shape filament to obtain a circular shape on the quiet Sun. In our observations, similar PRM also was found at the endpoint region of the forming filament. With magnetic helicity flux and connectivity-based density flux investigations, we found that the anti-clockwise PRM persistently injected negative helicity in the formation course of the filament. To some extent, both of these observations, including the present one, reveal that such PRMs may be not rare in the formation of solar filaments, and they might be considered as an observable tracer for the covert twist relaxations from below the photosphere to the low corona.
\par
In addition, the origin of filament plasma is another open question. Until now, several models have been proposed to answer this question. In the levitation models, cool dense filament plasma is thought to be lifted directly from the photosphere by emerging or reconnected field lines \citep{1994SoPh..155...69R}. In the injection models, cool plasma is proposed to be heated and injected into the corona $via$ successive chromospheric magnetic reconnection \citep{1999ApJ...520L..71W,2005ApJ...631L..93L,2016A&A...589A.114L,2016ApJ...831..123Z}. In the evaporation-condensation models, thermal non-equilibrium is believed to play a key role in the evaporation and condensation of filament dense plasma in the corona \citep{2005ApJ...635.1319K,2012ApJ...748L..26X}. Recently, \citet{2017ApJ...845...12K} conducted a three-dimensional MHD simulations for filament formation, in which anisotropic nonlinear thermal conduction and optically thin radiative cooling were considered. In their simulation, reconnection between sheared arcades leads not only to flux rope formation but also to condensation of filament plasma at the same time. In the present study, before the formation of the filament, many elongated chromospheric threads were created $via$ reconnection between the rising AFS and overlying field lines after flux emergence ended. Due to the persistent PRMs within their positive end region, these dispersive threads finally coalesced into the newborn filament. Such coalescence of dark threads is similar to that of \citet{2016SoPh..291.2373Z}, in which reconnection may also be inevitably involved. Thus one possible explanation for the origin of filament plasma in this event is that cool dense photospheric plasma initially lifted to chromospheric height by arch-shaped fields lines of the rising AFS \citep{1967SoPh....2..451B} and then a portion of them further collected in newborn sheared magnetic fields $via$ reconnection and radiative cooling processes in the formation course of the filament \citep{2017ApJ...845...12K}. Of course, contributions from other processes, such as evaporation or siphon effects, to the formation of filament plasma can not be ruled out in our observations. To fully understand this question, more new observations are required.


%
\begin{acks}
We thank the anonymous referees for their critical comments which helped improve the paper.
We also thank Professor Jun Zhang for constructive comments, Leping Li and Shuhong Yang for useful discussions. The data used here are courtesy of the NASA/SDO, the HMI and the AIA science teams.
This work is supported by the Natural Science Foundation of China under grants 11703084, 11633008,
11333007, and 11503081, and by the CAS programs "Light of West China" and "QYZDJ-SSW-SLH012", and by the grant associated with the Project of the Group for Innovation of Yunnan Province.
\end{acks}
\par


Disclosure of Potential Conflicts of Interest: The authors declare that they have no conflict of interest.
%
%
\bibliographystyle{spr-mp-sola}
\bibliography{bibtex}

%
%
%
\end{article}
\end{document}